\documentclass{emulateapj}

\def\kms{\ifmmode{\rm km\thinspace s^{-1}}\else km\thinspace s$^{-1}$\fi}
\def\vstar{V2154\,Cyg}

\shortauthors{Bright et al.}
\shorttitle{\vstar}

\begin{document}

\submitted{Accepted for publication in The Astrophysical Journal}

\title{Absolute dimensions of the F-type eclipsing binary V2154 Cygni}

\author{
Jane C.\ Bright\altaffilmark{1,2} and
Guillermo Torres\altaffilmark{1},
}

\altaffiltext{1}{Harvard-Smithsonian Center for Astrophysics, 60
  Garden St., Cambridge, MA 02138, USA; e-mail:
  gtorres@cfa.harvard.edu}

\altaffiltext{2}{Denison University, Granville, OH 43023, USA}

\begin{abstract}

We report spectroscopic observations of the 2.63 day, detached, F-type
main-sequence eclipsing binary \vstar. We use our observations
together with existing $uvby$ photometric measurements to derive
accurate absolute masses and radii for the stars good to better than
1.5\%. We obtain masses of $M_1 = 1.269 \pm 0.017~M_{\sun}$ and $M_2 =
0.7542 \pm 0.0059~M_{\sun}$, radii of $R_1 = 1.477 \pm 0.012~R_{\sun}$
and $R_2 = 0.7232 \pm 0.0091~R_{\sun}$, and effective temperatures of
$6770 \pm 150$~K and $5020 \pm 150$~K for the primary and secondary
stars, respectively. Both components appear to have their rotations
synchronized with the motion in the circular orbit. A comparison of
the properties of the primary with current stellar evolution models
gives good agreement for a metallicity of ${\rm [Fe/H]} = -0.17$,
which is consistent with photometric estimates, and an age of about
2.2~Gyr.  On the other hand, the K2 secondary is larger than predicted
for its mass by about 4\%. Similar discrepancies are known to exist
for other cool stars, and are generally ascribed to stellar
activity. The system is in fact an X-ray source, and we argue that the
main site of the activity is the secondary star. Indirect estimates
give a strength of about 1~kG for the surface magnetic field on that
star. A previously known close visual companion to \vstar\ is shown to
be physically bound, making the system a hierarchical triple.

\end{abstract}


\section{Introduction}
\label{sec:introduction}

\vstar\ (also known as HD~203839, HIP~105584, BD+47~3386, and
TYC~3594-1060-1; $V = 7.77$) is a 2.63 day eclipsing binary discovered
by the {\it Hipparcos\/} team \citep{Perryman:1997}, and found
independently in 1996 by \cite{Martin:2003} in the course of a search
for variable stars in the open cluster M39.  Light curves in the
$uvby$ Str\"omgren system were published by \cite{Rodriguez:2001}, but
the physical properties of the components were not derived by them
because spectroscopy was lacking. The only spectroscopic work we are
aware of are brief reports by \cite{Kurpinska:2000} listing
preliminary values for the velocity amplitudes, and by
\cite{Oblak:2004} giving preliminary masses and radii, though details
of those analyses are unavailable.  The very unequal depths of the
eclipses ($\sim$0.3 mag for the primary and $\sim$0.05 mag for the
secondary) suggest stars of rather different masses, making it an
interesting object for followup because of the increased leverage for
the comparison with stellar evolution models. This motivated us to
carry out our own high-resolution spectroscopic observations of this
star, which we report here.  \vstar\ is known from {\it Tycho-2}
observations to have a close, 0\farcs47 visual companion about two
magnitudes fainter than the binary \citep[$\Delta B_T = 2.18$~mag,
  $\Delta V_T = 2.15$~mag;][]{Fabricius:2000}. We show below that it
is physically associated, making \vstar\ a hierarchical triple system.

While the primary of the eclipsing pair is an early F star, the
secondary is a much smaller K star in the range where previous
observations have shown discrepancies with models \citep[see,
  e.g.,][]{Torres:2013}. The measured radii of such stars are
sometimes larger than predicted, and their temperatures cooler than
expected, both presumably due to the effects of magnetic activity
and/or spots \citep[e.g.,][]{Chabrier:2007, Morales:2010}.
\vstar\ therefore presents an opportunity to determine accurate
physical properties of the stars in a system with a mass ratio
significantly different from unity, and to investigate any
discrepancies with theory in connection with measures of stellar
activity.

The layout of our paper is as follows. Our new spectroscopic
observations are reported in Section~\ref{sec:spectroscopy}, followed
by a brief description in Section~\ref{sec:photometry} of the
\cite{Rodriguez:2001} photometric measurements we incorporate into our
analysis. The light curve fits are presented in
Section~\ref{sec:analysis}, along with consistency checks to support
the accuracy of the results. With the spectroscopic and photometric
parameters we then derive the physical properties of the system, given
in Section~\ref{sec:dimensions}, and compare them with current models
of stellar structure and stellar evolution
(Section~\ref{sec:models}). We discuss the results in the context of
available activity measurements in Section~\ref{sec:discussion}, and
conclude with some final thoughts in Section~\ref{sec:conclusions}.

\section{Spectroscopic observations and analysis}
\label{sec:spectroscopy}

\vstar\ was placed on our spectroscopic program in October of 2001,
and observed through June of 2007 with two nearly identical echelle
instruments \citep[Digital Speedometer;][]{Latham:1992} on the 1.5\,m
telescope at the Oak Ridge Observatory in the town of Harvard (MA),
and on the 1.5\,m Tillinghast reflector at the Fred L.\ Whipple
Observatory on Mount Hopkins (AZ). Both instruments (now
decommissioned) used intensified photon-counting Reticon detectors
providing spectral coverage in a single echelle order 45~\AA\ wide
centered on the \ion{Mg}{1}\,b triplet at 5187~\AA. The resolving
power delivered by these spectrographs was $R \approx 35,\!000$, and
the signal-to-noise ratios achieved for the 80 usable observations of
\vstar\ range from about 20 to 67 per resolution element of 8.5~\kms.
Wavelength solutions were carried out by means of exposures of a
thorium-argon lamp taken before and after each science exposure, and
reductions were performed with a custom pipeline. Observations of the
evening and morning twilight sky were used to place the observations
from the two instruments on the same velocity system and to monitor
instrumental drifts \citep{Latham:1992}.

Visual inspection of one-dimensional cross-correlation functions for
each of our spectra indicated the presence of a star much fainter than
the primary that we initially assumed was the secondary in \vstar.
However, subsequent analysis with the two-dimensional
cross-correlation algorithm TODCOR \citep{Zucker:1994} showed those
faint lines to be stationary, while a third set of even weaker lines
was noticed that moved in phase with the orbital period. This is
therefore the secondary in the eclipsing pair, and the stationary
lines correspond to the visual companion mentioned in the
Introduction, as we show later, which falls within the 1\arcsec\ slit
of the spectrograph. Consequently, for the final velocity measurements
we used an extension of TODCOR to three dimensions \citep[referred to
  here as TRICOR;][]{Zucker:1995} that uses three different templates,
one for each star. In the following we refer to the binary components
as stars 1 and 2, and to the tertiary as star 3.  The templates were
selected from a large library of synthetic spectra based on model
atmospheres by R.\ L.\ Kurucz \citep[see][]{Nordstrom:1994,
  Latham:2002}, computed for a range of temperatures ($T_{\rm eff}$),
surface gravities ($\log g$), rotational broadenings ($v \sin i$, when
seen in projection), and metallicities ([m/H]).

We selected the optimum parameters for the templates as follows,
adopting solar metallicity throughout. For the primary star we ran a
grid of one-dimensional cross-correlations against synthetic spectra
over a wide range of temperatures and $v \sin i$ values
\citep[see][]{Torres:2002}, for a fixed $\log g$ of 4.0 that is
sufficiently close to our final estimate presented later.  The best
match, as measured by the cross-correlation coefficient averaged over
all exposures, was obtained for interpolated values of $T_{\rm eff} =
6770 \pm 150$~K and $v \sin i = 26 \pm 2$~\kms.  The secondary and
tertiary stars are faint enough (by factors of 25 and 9, respectively;
see below) that they do not affect these results significantly.  For
the secondary the optimal $v \sin i$ from grids of TRICOR correlations
was $12 \pm 2$~\kms. However, due to its faintness we were unable to
establish its temperature from the spectra themselves, so we relied on
results from the light curve analysis described later in
Section~\ref{sec:analysis}. The central surface brightness ratio $J$
provides an accurate measure of the temperature ratio between stars 1
and 2. Using the primary temperature from above, the $J$ value for the
$y$ band, and the visual flux calibration by \cite{Popper:1980}, we
obtained $T_{\rm eff} = 5020 \pm 150$~K. The surface gravity was
adopted as $\log g = 4.5$, appropriate for a main-sequence star of
this temperature. For the tertiary we again adopted $\log g = 4.5$,
and grids of correlations with TRICOR for a range of temperatures
indicated a preference for a value of 5500~K, to which we assign a
conservative uncertainty of 200~K.  Similar correlation grids varying
$v \sin i$ indicated no measurable line broadening for the tertiary,
so we adopted $v \sin i = 0$~kms, with an estimated upper limit of
2~\kms.

Radial velocities were then measured with TRICOR using values for the
template parameters ($T_{\rm eff}$, $v \sin i$) in our library nearest
to those given above: 6750~K and 25~\kms\ for the primary, 5000~K and
12~\kms\ for the secondary, and 5500~K and 0~\kms\ for the
tertiary. The light ratios we determined from our spectra are
$L_2/L_1 = 0.036 \pm 0.004$ and $L_3/L_1 = 0.108 \pm
0.012$, corresponding to the mean wavelength of our observations
(5187~\AA).

\begin{deluxetable*}{lccccccc}
\tablewidth{0pc}
\tablecaption{Heliocentric radial velocity measurements of \vstar.
 \label{tab:rvs}}
\tablehead{
\colhead{HJD} &
\colhead{$RV_1$} &
\colhead{$\sigma_1$} &
\colhead{$RV_2$} &
\colhead{$\sigma_2$} &
\colhead{$RV_3$} &
\colhead{$\sigma_3$} &
\colhead{Orbital}
\\
\colhead{(2,400,000$+$)} &
\colhead{(\kms)} &
\colhead{(\kms)} &
\colhead{(\kms)} &
\colhead{(\kms)} &
\colhead{(\kms)} &
\colhead{(\kms)} &
\colhead{phase}
}
\startdata
 51874.5314 &  \phs36.96 &   1.11 &  \phn$-$20.74 &   9.28 &   17.32 &   1.85 &  0.4586 \\
 52109.6581 &   $-$41.58 &   0.67 &  \phs128.03   &   5.55 &   19.35 &   1.11 &  0.8387 \\
 52123.6422 &  \phs79.58 &   0.65 &  \phn$-$85.40 &   5.42 &   19.62 &   1.08 &  0.1546 \\
 52130.5621 &   $-$52.47 &   0.75 &  \phs135.05   &   6.22 &   18.83 &   1.24 &  0.7851 \\
 52151.5379 &   $-$53.69 &   0.69 &  \phs143.48   &   5.73 &   19.73 &   1.14 &  0.7588
\enddata
\tablecomments{Orbital phases are computed from the reference time of
  primary eclipse given in Section~\ref{sec:analysis}. This table is
  available in its entirety in machine-readable form.}
\end{deluxetable*}

Because our spectra are only 45~\AA\ wide, systematic errors in the
velocities can result from lines shifting in and out of this window as
a function of orbital phase \citep[see][]{Latham:1996}. To estimate
this effect we followed a procedure similar to that of
\cite{Torres:1997} and created artificial triple-lined spectra based
on our adopted templates, which we then processed with TRICOR in the
same way as the real spectra. A comparison of the input and output
velocities showed a phase-dependent pattern with maximum shifts of
about 0.2~\kms\ for the primary, 6~\kms\ for the secondary, and
1.2~\kms\ for the tertiary.  We applied these shifts as corrections to
the individual raw velocities, and the final measurements including
all corrections are listed in Table~\ref{tab:rvs}, along with their
uncertainties. The velocities of the third star appear constant within
their uncertainties, and have a mean of $+19.31 \pm
0.13$~\kms\ (weighted average). A similar correction for systematic
errors was applied to the light ratios, and is already included in the
values reported above.

\begin{figure}
\epsscale{1.15} \plotone{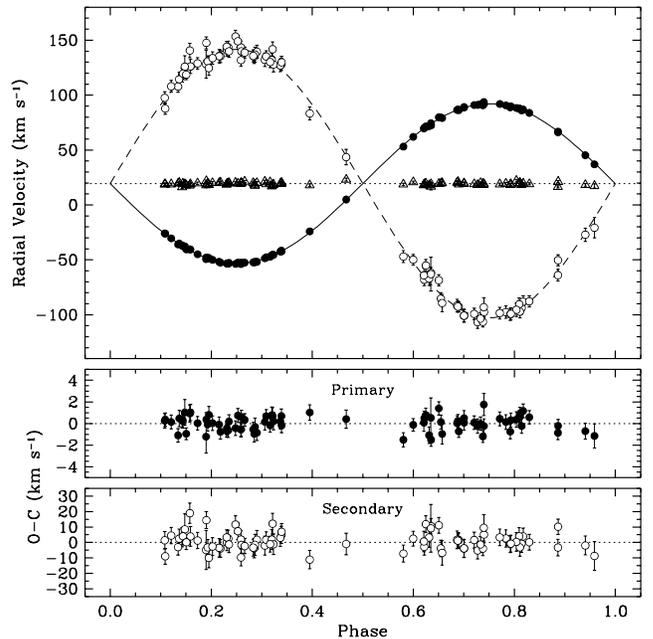}

\figcaption{Radial-velocity measurements of \vstar\ along with our
  orbital fit. The primary is represented with filled circles, the
  secondary with open circles, and the tertiary with open triangles.
  The dotted line marks the center-of-mass velocity $\gamma$, and
  phases are computed from the reference time of primary
  eclipse. Residuals are shown at the bottom. \label{fig:rvs} }
\end{figure}

A weighted least-squares orbital fit to the primary and secondary
velocities gives the elements and derived quantities presented in
Table~\ref{tab:specorbit}, where a circular orbit has been
assumed. Tests allowing for eccentricity gave results consistent with
zero, in agreement with similar experiments below based on the light
curves. Initial solutions in which we included a possible systematic
offset between the primary and secondary velocities, as may arise,
e.g., from template mismatch, also gave a value consistent with zero.
The observations and orbital fit are shown in Figure~\ref{fig:rvs}.
The tertiary velocities, represented with triangles, are seen to be
very close to the center-of-mass velocity, supporting the physical
association.

\begin{deluxetable}{lc}
\tablewidth{0pc}
\tablecaption{Spectroscopic orbital solution for \vstar.\label{tab:specorbit}}
\tablehead{
\colhead{
\hfil~~~~~~~~~~~~~Parameter~~~~~~~~~~~~~~} & \colhead{Value}}
\startdata
\multicolumn{2}{c}{Orbital elements\hfil}                                        \\
\noalign{\vskip 2pt}
\noalign{\hrule}
\noalign{\vskip 3pt}
~~~~$P$ (days)\tablenotemark{a}                &  2.6306359~$\pm$~0.0000039       \\
~~~~$T_{\rm max}$ (HJD$-$2,400,000)\tablenotemark{a}   & 52973.58847~$\pm$~0.00091\phm{2222} \\
~~~~$\gamma$ (\kms)                            & +19.408~$\pm$~0.076\phs\phn   \\
~~~~$K_1$ (\kms)                        &   72.699~$\pm$~0.092\phn \\
~~~~$K_2$ (\kms)                        &   122.298~$\pm$~0.723\phn\phn \\
~~~~$e$                                        & 0.0 (fixed)                 \\
\noalign{\vskip 2pt}
\noalign{\hrule}
\noalign{\vskip 3pt}
\multicolumn{2}{c}{Derived quantities\hfil} \\
\noalign{\vskip 2pt}
\noalign{\hrule}
\noalign{\vskip 3pt}
~~~~$M_1\sin^3 i$ ($M_{\sun}$)\tablenotemark{b}   &  1.268~$\pm$~0.017          \\
~~~~$M_2\sin^3 i$ ($M_{\sun}$)\tablenotemark{b}   &  0.7535~$\pm$~0.0059          \\
~~~~$q\equiv M_2/M_1$            &  0.5944~$\pm$~0.0036          \\
~~~~$a_1\sin i$ (10$^6$ km)             &  2.6298~$\pm$~0.0033          \\
~~~~$a_2\sin i$ (10$^6$ km)             &  4.424~$\pm$~0.026          \\
~~~~$a \sin i$ ($R_{\sun}$)\tablenotemark{b}   &  10.140~$\pm$~0.038\phn          \\
\noalign{\vskip 2pt}
\noalign{\hrule}
\noalign{\vskip 3pt}
\multicolumn{2}{c}{Other quantities pertaining to the fit\hfil}              \\
\noalign{\vskip 2pt}
\noalign{\hrule}
\noalign{\vskip 3pt}
~~~~$N_{\rm obs}$                              & 80                          \\
~~~~Time span (days)                           &  2381.4                     \\
~~~~Time span (cycles)                         &  905.3                     \\
~~~~$\sigma_1$ (\kms)                   & 0.69                         \\
~~~~$\sigma_2$ (\kms)                   & 5.74                          \\
~~~~$\sigma_3$ (\kms)                    & 1.21                          
\enddata
\tablenotetext{a}{Time of maximum primary velocity.}
\tablenotetext{b}{Based on physical constants recommended by 2015 IAU Resolution B3 \citep{Prsa:2016}.}
\end{deluxetable}

\section{Photometric observations}
\label{sec:photometry}

The light curves used for our analysis are those published by
\cite{Rodriguez:2001}\footnote{We note, incidentally, that the
  heliocentric Julian dates in the online electronic files should be
  corrected by subtracting exactly 790 days.}, and were obtained
between July and November of 1998 with the 0.9\,m telescope at the
Sierra Nevada Observatory (Spain).  The 852 observations were made on
28 nights using $uvby$ filters, with HD\,204626 (\ion{A0}{3}) as the
comparison star and HD\,204977 (\ion{B9}{5}) as the check star. The
standard deviations of the difference in magnitude between the
comparison and check stars, which may be taken as an indication of the
precision of the observations, were 0.0085, 0.0035, 0.0032, and
0.0043~mag for $u$, $v$, $b$, and $y$, respectively.

\section{Light curve analysis}
\label{sec:analysis}

For the analysis of the light curves of this well-detached system we
have adopted the Nelson-Davis-Etzel model \citep{Popper:1981,
  Etzel:1981}, as implemented in the JKTEBOP code\footnote{{\url
    http://www.astro.keele.ac.uk/jkt/codes/jktebop.html}}
\citep{Southworth:2013}. The free parameters of the fit are the period
$P$ and reference epoch of primary minimum $T_{\rm min}$, the central
surface brightness ratio $J \equiv J_2/J_1$, the sum of the relative
radii $r_1+r_2$ normalized to the semimajor axis, the radius ratio $k
\equiv r_2/r_1$, the inclination angle $i$, and a magnitude zero point
$m_0$. Because of the presence of the third star in the aperture we
also included the third light parameter $L_3$ (fractional brightness
of star 3 divided by the total light, at phase 0.25 from primary
eclipse). The mass ratio was held fixed at the spectroscopic value ($q
= 0.5944$).  Linear limb-darkening coefficients ($u_1$, $u_2$) were
interpolated from the tables by \cite{Claret:2000} using the JKTLD
code\footnote{\url{http://www.astro.keele.ac.uk/jkt/codes/jktld.html}}
\citep{Southworth:2008}, and gravity-darkening coefficients ($y_1$,
$y_2$) were taken from the tabulations by \cite{Claret:2011} for the
properties of the primary and secondary given earlier. Experiments
with quadratic limb-darkening gave no improvement, so the linear law
was used throughout.  Initial fits that included the eccentricity as
an additional free parameter indicated a value that was not
significantly different from zero, consistent with the spectroscopic
evidence, so the orbit was assumed to be circular.

\begin{deluxetable*}{lcccc}
\tablewidth{0pc}
\tablecaption{Light curve solutions for \vstar.\label{tab:LCfits1}}
\tablehead{
\colhead{
\hfil~~Parameter~~} &
\colhead{$u$} &
\colhead{$v$} &
\colhead{$b$} &
\colhead{$y$}
}
\startdata
$P$ (days)                       &  2.630607 (+14/$-$19)    &  2.6306290 (+61/$-$75)   &  2.6306305 (+56/$-$59)   &  2.6306316 (+74/$-$81) \\
$T_{\rm min}$ (HJD$-$2,400,000)  &  51048.61797 (+28/$-$22) &  51048.61815 (+12/$-$14) &  51048.61814 (+11/$-$14) &  51048.61808 (+15/$-$12) \\
$r_1+r_2$          &  0.2169 (+26/$-$17)      &  0.2172 (+12/$-$14)      &  0.2167 (+13/$-$15)      &  0.2163 (+13/$-$14) \\
$k \equiv r_2/r_1$ &  0.492 (+17/$-$14)       &  0.486 (+18/$-$9)        &  0.492 (+15/$-$14)       &  0.473 (+24/$-$4)  \\
$i$ (deg)                        &  88.76 (+32/$-$79)       &  88.39 (+61/$-$33)       &  88.57 (+43/$-$48)       &  87.79 (+82/$-$6)  \\
$J$                              &  0.120 (+11/$-$9)        &  0.1267 (+84/$-$58)      &  0.193 (+11/$-$9)        &  0.246 (+16/$-$11) \\
$L_3$                            &  0.087 (+35/$-$59)       &  0.054 (+51/$-$29)       &  0.093 (+34/$-$47)       &  0.028 (+87/$-$1)  \\
$m_0$ (mag)                      &  0.33069 (+45/$-$45)     &  0.69957 (+40/$-$32)     &  0.44492 (+37/$-$31)     &  0.20558 (+44/$-$29) \\
\noalign{\vskip 2pt}
\noalign{\hrule}
\noalign{\vskip 3pt}
\multicolumn{5}{c}{Derived quantities} \\
\noalign{\vskip 2pt}
\noalign{\hrule}
\noalign{\vskip 3pt}
$r_1$                     &   0.1454 (+24/$-$22)  &  0.1462 (+13/$-$23)   &  0.1453 (+20/$-$23)  &  0.1469 (+11/$-$31) \\
$r_2$                     &   0.0715 (+16/$-$12)  &  0.0710 (+15/$-$8)    &  0.0714 (+12/$-$13)  &  0.0695 (+21/$-$2)  \\
$L_2/L_1$          &   0.0264 (+21/$-$23)  &  0.0280 (+22/$-$10)   &  0.0435 (+22/$-$24)  &  0.0513 (+60/$-$5)  \\
$\sigma$ (mmag)                  &   8.56              &  3.58               &  3.28              &  3.79             \\
\noalign{\vskip 2pt}
\noalign{\hrule}
\noalign{\vskip 3pt}
\multicolumn{5}{c}{Adopted limb-darkening and gravity-darkening coefficients \citep{Claret:2000, Claret:2011}} \\
\noalign{\vskip 2pt}
\noalign{\hrule}
\noalign{\vskip 3pt}
$u_1$                     &  0.722           &  0.748           &  0.696           &  0.615 \\
$u_2$                     &  0.929           &  0.892           &  0.854           &  0.768 \\
$y_1$                     &  0.393           &  0.354           &  0.305           &  0.260 \\
$y_2$                     &  1.157           &  0.892           &  0.672           &  0.581 
\enddata

\tablecomments{Uncertainties from the residual permutation procedure
  are given in parentheses in units of the last significant place
  (upper and lower error bars).}

\end{deluxetable*}

Separate solutions for each of the $uvby$ bands are presented in
Table~\ref{tab:LCfits1}. As the errors provided by JKTEBOP are
strictly internal and do not capture systematic components that may
result, e.g., from red noise, the uncertainties given in the table
were computed with the residual permutation (``prayer bead'') method,
as follows. We shifted the residuals from the original fits by an
arbitrary number of time indices (with wraparound), and added them
back into the computed curves to create artificial data sets that
preserve any time-correlated noise that might be present in the
original data. We generated 500 such data sets for each of the
passbands and fitted them with JKTEBOP.  In each solution we
simultaneously perturbed all of the quantities that were initially
held fixed. We did this by adding Gaussian noise to the mass ratio
corresponding to its measured error ($\sigma_q = 0.0036$), and
Gaussian noise with $\sigma = 0.1$ to the limb-darkening and
gravity-darkening coefficients. The standard deviations of the
resulting distributions for each parameter were adopted as the
uncertainties for the light curve elements.

\begin{deluxetable}{lc}
\tablewidth{0pc}
\tablecaption{Adopted ephemeris and geometric light curve elements for \vstar.\label{tab:LCfits2}}
\tablehead{
\colhead{~~~~Parameter~~~~} &
\colhead{Value}
}
\startdata
$r_1+r_2$          &  0.21696~$\pm$~0.00087 \\
$k \equiv r_2/r_1$ &  0.4895~$\pm$~0.0083 \\
$i$ (deg)                        &  88.55~$\pm$~0.28\phn \\
$r_1$                     &   0.1457~$\pm$~0.0010 \\
$r_2$                     &   0.07129~$\pm$~0.00060 \\
$P$ (days)                       &  2.6306303~$\pm$~0.0000038 \\
$T_{\rm min}$ (HJD$-$2,400,000)  &  51048.618122~$\pm$~0.000075\phm{2222}
\enddata
\end{deluxetable}

\setlength{\tabcolsep}{4pt}
\begin{deluxetable}{cccc}
\tablewidth{0pc}
\tablecaption{Adopted wavelength-dependent light curve elements \\ for \vstar.\label{tab:LCfits3}}
\tablehead{
\colhead{$\lambda$} &
\colhead{$J$} &
\colhead{$L_3$} &
\colhead{$L_2/L_1$}
}
\startdata
$u$ &  0.119 (+12/$-$10)  & 0.075 (+22/$-$13) & 0.0259 (+19/$-$16) \\
$v$ &  0.1272 (+86/$-$82) & 0.069 (+22/$-$22) & 0.02851 (+75/$-$67) \\
$b$ &  0.193 (+11/$-$11)  & 0.086 (+22/$-$21) & 0.04311 (+81/$-$75) \\
$y$ &  0.250 (+12/$-$14)  & 0.101 (+21/$-$17) & 0.0560 (+13/$-$10)
\enddata
\tablecomments{Uncertainties from the residual permutation procedure
  are given in parentheses in units of the last significant place
  (upper and lower error bars).}
\end{deluxetable}
\setlength{\tabcolsep}{6pt}

The results from the four passbands are fairly consistent within their
uncertainties, with a few exceptions: (1) The ephemeris ($P$, $T_{\rm
  min}$) seems rather different for the $u$ band, which is the fit
with the largest scatter. The fact that the $uvby$ measurements are
simultaneous indicates this is almost certainly due to systematic
errors affecting $u$ that are not uncommon. (2) The geometric
parameters (most notably $k$ and $i$, and to a lesser extent
$r_1+r_2$) seem systematically different for the $y$ band. Several
features of that fit make us suspicious of these quantities, and of
$L_3$ as well. In particular, $L_3$ is significantly lower than in the
other bands, which runs counter to expectations given that the third
star is cooler (redder) than the primary, and so its flux contribution
ought to be larger in $y$, not smaller. Third light is always strongly
(and positively) correlated with the inclination angle and with $k$ in
this case, and indeed we see that both $i$ and $k$ are also low. Grids
of JKTEBOP solutions over a range of fixed values of $k$ show that for
all $k$ values the radius sum in the $y$ band is always considerably
smaller than in the other three bands, which agree well among each
other. Finally, we note that the $y$-band error bars for $k$, $i$, and
$L_3$ are all highly asymmetric (always much larger in the direction
toward the average of the $uvb$ results), which is not the case in the
other bands.  These features are symptomatic of strong degeneracies in
$y$ that make the results highly prone to biases. We have therefore
chosen not to rely on the geometric parameters from the $y$ band.

\begin{figure}
\epsscale{1.15}
\plotone{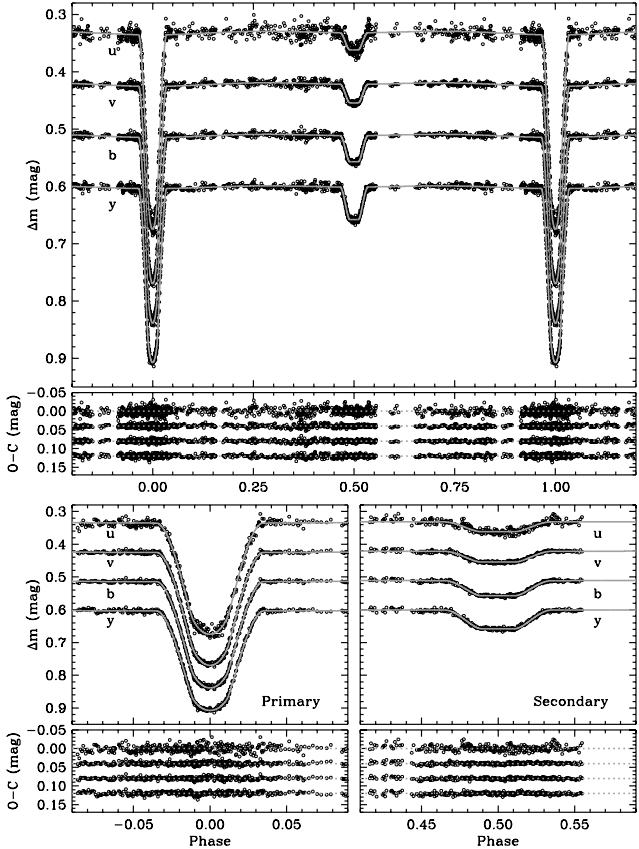}

\figcaption{\emph{Top:} Light curves of \vstar\ along with our model
  fits. The $v$, $b$, and $y$ bands as well as their residuals have
  been shifted vertically relative to the $u$ band for
  clarity. \emph{Bottom:} Enlargements around the primary and
  secondary minima.\label{fig:LC} }

\end{figure}

Weighted averages of the photometric period and epoch (excluding the
$u$ band) and of the geometric parameters (excluding the $y$ band) are
given in Table~\ref{tab:LCfits2}. The photometric period agrees well
with the spectroscopic one, within the errors. The final solutions for
the wavelength-dependent quantities were carried out by holding the
ephemeris and geometry fixed to these values, and the results are
collected in Table~\ref{tab:LCfits3}. We illustrate these final fits
in Figure~\ref{fig:LC}, where the secondary eclipse is seen to be
total.


\subsection{Consistency checks}
\label{sec:consistency}

The spectroscopic light ratios reported in
Section~\ref{sec:spectroscopy} ($L_2/L_1$ and $L_3/L_1$), which are
independent of the light curve analysis above, offer an opportunity to
test the accuracy of the light curve solutions. For the necessary flux
transformation between the 5187~\AA\ spectral window and the slightly
redder Str\"omgren $y$ band (5470~\AA) we used synthetic spectra from
the PHOENIX library by \cite{Husser:2013}, along with our adopted
effective temperatures and surface gravities from
Section~\ref{sec:spectroscopy}, integrating the model fluxes over both
passbands. An additional quantity that is needed to properly scale the
spectral energy distributions is the radius ratio.

\begin{figure}
\epsscale{1.15}
\plotone{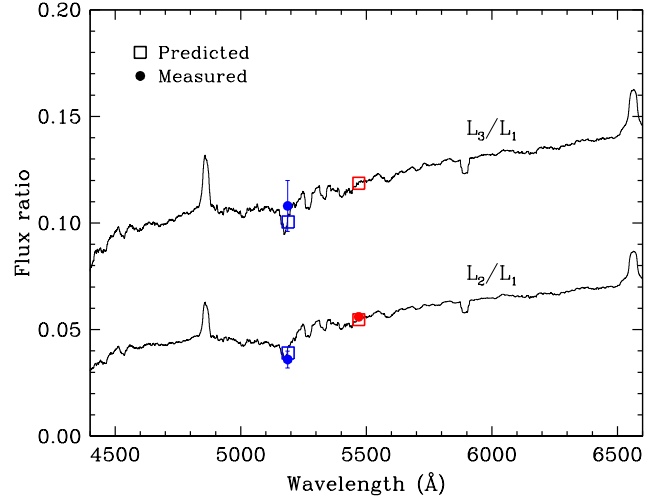}

\figcaption{Calculated flux ratios $L_2/L_1$ and $L_3/L_1$
  as a function of wavelength, from synthetic spectra by
  \cite{Husser:2013} for the adopted temperatures and surface
  gravities of the three stars. The bottom curve has no free
  parameters and matches the $y$-band ratio from the light curve as
  well as the spectroscopic 5187~\AA\ ratio. The top curve was
  adjusted (using $R_3/R_1 = 0.56$) to match the $y$-band flux ratio,
  and is seen to also match the spectroscopic ratio
  well.\label{fig:bands} }

\end{figure}

As a sanity check we first used these spectra coupled with our
measured radius ratio of $k = 0.4895$ to calculate the $y$-band light
ratio between the primary and secondary, and obtained $L_2/L_1 =
0.055$, in good agreement with our light curve value. The flux ratio
we then infer at 5187~\AA\ based on the same parameters is 0.039,
which is consistent with the spectroscopic measurement of $0.036 \pm
0.004$ (see Figure~\ref{fig:bands}).

The scaling of the energy distributions of the tertiary and primary
components requires knowledge of the radius ratio between those two
stars, which our observations do not provide. We estimated it as
follows. With our $y$-band light curve results from
Table~\ref{tab:LCfits3} ($L_2/L_1$ and $L_3$) we calculated $L_3/L_1 =
L_3 (1 + L_2/L_1)/(1 - L_3) = 0.119$. We then used the PHOENIX
synthetic spectra and varied the radius ratio until we reproduced this
value of $L_3/L_1$, which occurred for $R_3/R_1 = 0.56$. With the
scaling set in this way, the predicted flux ratio at 5187~\AA\ between
the tertiary and primary is 0.100, which again agrees with the
spectroscopically measured ratio of $0.108 \pm 0.012$, as illustrated
in Figure~\ref{fig:bands}.

These consistency checks between the spectroscopy and the photometry
are an indication that the light curve fits are largely free from
biases, and support the accuracy of the geometric elements used in the
next section to derive the physical properties of the stars.

\section{Absolute dimensions}
\label{sec:dimensions}

The absolute masses and radii of \vstar\ are listed in
Table~\ref{tab:dimensions}. The relative uncertainties are smaller
than 1.5\% for both components. The combined out-of-eclipse magnitudes
of the system from \cite{Rodriguez:2001} and our fitted light ratios
and third-light values enable us to deconvolve the light of the
components. For the primary star we obtained the Str\"omgren indices
$b-y = 0.243 \pm 0.035$, $m_1 = 0.139 \pm 0.063$, and $c_1 = 0.528 \pm
0.063$, along with $\beta = 2.691$. With these and the calibrations of
\cite{Crawford:1975} we infer negligible reddening for the system
(consistent with its small distance; see below), and an estimated
photometric metallicity of ${\rm [Fe/H]} = -0.12$.  Photometric
estimates of the temperatures may be obtained from the $b-y$ index of
the primary and the corresponding value for the secondary of $0.527
\pm 0.046$. The color/temperature calibration of
\cite{Casagrande:2010} leads to values of $6840 \pm 200$~K and $5050
\pm 260$~K that are in good agreement with the spectroscopic values
adopted in Section~\ref{sec:spectroscopy}. The deconvolved color of
the third star ($b-y = 0.45 \pm 0.38$) is too uncertain to be useful,
though the inferred temperature of $5500 \pm 870$~K again matches the
value from Section~\ref{sec:spectroscopy}. The spectral types
corresponding to the adopted temperatures are F2, K2, and G8 for the
primary, secondary, and tertiary, respectively.

Additional quantities listed in Table~\ref{tab:dimensions} include the
luminosities, absolute magnitudes, and the distance ($90 \pm 9$~pc),
which makes use of the bolometric corrections by \citep{Flower:1996}.
The corresponding parallax, $11.2 \pm 1.1$~mas, is not far from the
trigonometric values listed in the {\it Hipparcos\/} catalog
($\pi_{\rm HIP} = 11.77 \pm 0.59$~mas) and in the first data release
of {\it Gaia\/} \citep[$\pi_{\rm Gaia} = 13.35 \pm
  0.82$~mas;][]{Brown:2016}.  Our measured projected rotational
velocities are also quite close to the expected synchronous values
($v_{\rm sync} \sin i$).

As noted earlier, the third star was angularly resolved by the {\it
  Tycho-2\/} experiment at a separation of 0\farcs47 and a measured
position angle of 59\arcdeg, at the mean epoch 1991.25.  Subsequent
astrometric measurements by a number of authors indicate a gradual
decrease in the angular separation to 0\farcs25 in 2010
\citep{Horch:2010}, with no significant change in the position angle.
This is inconsistent with being the result of a chance alignment with
a background star, as the binary's fairly large proper motion of
113~mas~yr$^{-1}$ measured by {\it Gaia\/} would have carried the
companion 2\arcsec\ away in that interval. The direction of motion
would suggest a high inclined orbit, or possibly even an edge-on
orientation. At our measured 90~pc distance the 0\farcs47 separation
implies a semimajor axis of roughly 42~au and an orbital period of
$\sim$160~yr.

\begin{deluxetable}{lcc}
\tablewidth{0pt}
\tablecaption{Physical properties of \vstar.\label{tab:dimensions}}
\tablehead{
\colhead{~~~~~~~~~Parameter~~~~~~~~~} &
\colhead{Primary} &
\colhead{Secondary}
}
\startdata
Mass ($M_{\sun}$)\dotfill                          & $1.269 \pm 0.017$          &  $0.7542 \pm 0.0059$    \\ 
Radius ($R_{\sun}$)\dotfill                        & $1.477 \pm 0.012$          &   $0.7232 \pm 0.0091$   \\
$\log g$ (cgs)\dotfill                             & $4.2028 \pm 0.0089$        &   $4.597 \pm 0.012$     \\ 
Temperature (K)\dotfill                            & $6770 \pm 150$\phn         &   $5020 \pm 150$\phn    \\
$\log L/L_{\sun}$\dotfill                          & $0.616 \pm 0.039$          &  $-0.523 \pm 0.039$\phs \\
$BC_{\rm V}$ (mag)\tablenotemark{a}\dotfill        & $-0.02 \pm 0.10$\phs       &    $-0.30 \pm 0.11$\phs \\
$M_{\rm bol}$ (mag)\tablenotemark{b}\dotfill       & $3.192 \pm 0.098$          &  $6.041 \pm 0.097$      \\
$M_V$ (mag)\dotfill                                & $3.17 \pm 0.14$            &  $6.34 \pm 0.17$        \\
$m-M$ (mag)\dotfill                                & \multicolumn{2}{c}{$4.78 \pm 0.21$}                  \\
Distance (pc)\tablenotemark{c}\dotfill             & \multicolumn{2}{c}{$90 \pm 9$\phn}                   \\ 
Parallax (mas)\tablenotemark{c}\dotfill            & \multicolumn{2}{c}{$11.2 \pm 1.1$\phn}               \\
$v_{\rm sync} \sin i$ (\kms)\dotfill               & $28.4 \pm 0.2$\phn         &  $13.9 \pm 0.2$\phn     \\
$v \sin i$ (\kms)\tablenotemark{d}\dotfill         &   $26 \pm 2$\phn           &    $12 \pm 2$\phn
\enddata
\tablenotetext{a}{Bolometric corrections from \cite{Flower:1996}, with
  a contribution of 0.10 mag added in quadrature to the uncertainty
  from the temperatures.}
\tablenotetext{b}{Uses $M_{\rm bol}^{\sun} = 4.732$ for consistency
  with the adopted table of bolometric
  corrections \citep[see][]{Torres:2010}.}
\tablenotetext{c}{Relies on the luminosities, the apparent
  magnitude of \vstar\ out of eclipse \citep[$V = 7.773 \pm
    0.008$;][]{Rodriguez:2001}, and bolometric corrections.}
\tablenotetext{d}{Measured value.}
\end{deluxetable}

\section{Comparison with stellar evolution models}
\label{sec:models}

\begin{figure}
\epsscale{1.15}
\plotone{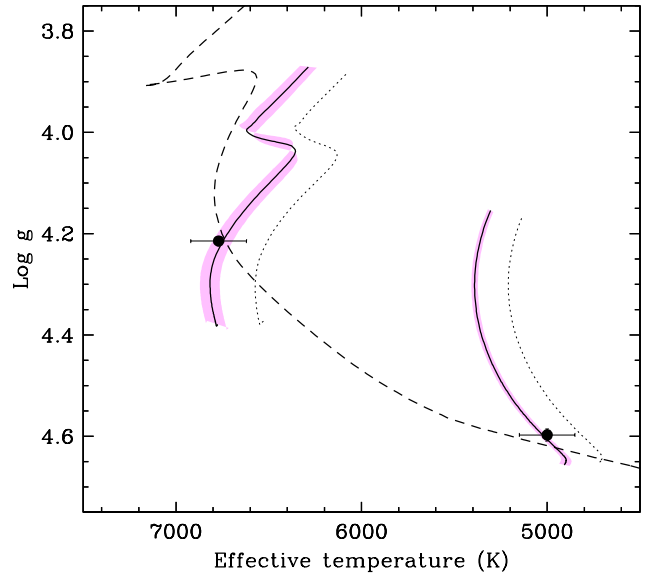}

\figcaption[]{Measurements for \vstar\ in the $\log g$ vs.\ $T_{\rm
    eff}$ diagram compared with evolutionary tracks (solid lines) from
  the MIST series \citep{Choi:2016} for a metallicity of ${\rm [Fe/H]}
  = -0.17$ that best matches the observations. The shaded areas around
  the solid primary and secondary tracks give an indication of the
  uncertainty in the measured masses. Evolutionary tracks for solar
  metallicity are shown with dotted lines, for reference. The dashed
  line represents a 2.2~Gyr isochrones that provides the best fit to
  the radius of the primary (see
  Figure~\ref{fig:mistmassradius}).\label{fig:mistlogg}}
\end{figure}

The accurate properties for \vstar\ are compared with predictions from
current stellar evolution theory in Figure~\ref{fig:mistlogg}. The
evolutionary tracks for the measured masses of the components were
taken from the grid of MESA Isochrones and Stellar Tracks
\citep[MIST;][]{Choi:2016}, which is based on the Modules for
Experiments in Stellar Astrophysics package
\citep[MESA;][]{Paxton:2011, Paxton:2013, Paxton:2015}. The
metallicity in the models was adjusted to ${\rm [Fe/H]} = -0.17$ to
provide the best fit to the temperatures of the stars. This
composition is not far from the photometric estimate reported earlier.
The shaded areas in the figure indicate the uncertainty in the
location of the tracks that comes from the errors in the measured
masses. Solar-metallicity tracks are shown with dotted lines for
reference.  The age that best matches the radius of the primary is
2.2~Gyr (see below). An isochrone for this age is shown with a dashed
line. The primary star is seen to be almost halfway through its
main-sequence phase.

At this relatively old age it is not surprising that we found the
components' rotation to be synchronized with the motion in a circular
orbit, as the theoretically expected timescales for synchronization
and orbit circularization are $\sim$1~Myr and $\sim$200~Myr,
respectively \citep[e.g.,][]{Hilditch:2001}.

\begin{figure}
\epsscale{1.15}
\plotone{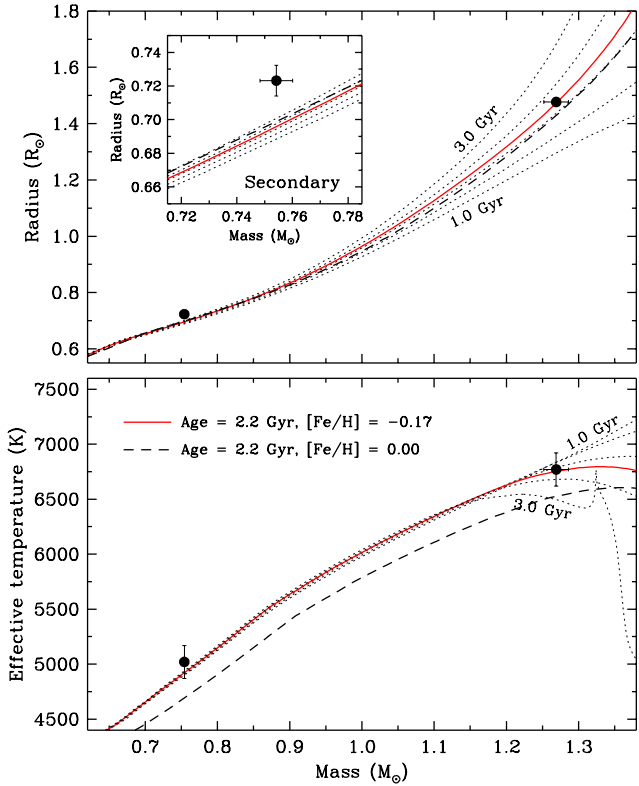}
\figcaption[]{Measured masses, radii, and temperatures of
  \vstar\ compared against model isochrones from the MIST series
  \citep{Choi:2016} for the same best-fit metallicity of ${\rm [Fe/H]}
  = -0.17$ as in Figure~\ref{fig:mistlogg}. The solid red line
  corresponds to an age of 2.2~Gyr that matches the size of the
  primary star, and dotted lines represent ages from 1.0 to 3.0~Gyr in
  steps of 0.5~Gyr at this composition. A 2.2~Gyr isochrone for solar
  metallicity is shown by the dashed line. The inset in the top panel
  shows an enlargement around the secondary
  star, revealing it to be ``inflated''. \label{fig:mistmassradius}}
\end{figure}

The radii and temperatures are shown separately as a function of mass
in Figure~\ref{fig:mistmassradius}, in which the solid line represents
the 2.2~Gyr isochrone for ${\rm [Fe/H]} = -0.17$ that reproduces the
measured radius of the primary star at its measured mass. A solar
metallicity isochrone for the same age is shown with the dashed
line. The secondary star is seen to be larger than predicted for its
mass by almost 4\%, corresponding to a nearly $\sim$3$\sigma$
discrepancy.  Similar deviations from theory are known to be present
in other stars with convective envelopes, and are usually attributed
to the effects of stellar activity \citep[see, e.g.][]{Popper:1997,
  Torres:2013}.  The bottom panel of the figure shows that the
temperatures of the two components are consistent with the theoretical
values for their mass within the errors. This is somewhat unexpected
for the secondary, as stellar activity typically causes both ``radius
inflation'' and ``temperature suppression'', though the latter effect
is smaller and not as easy to detect.

Aside from the brightness measurements and our spectroscopic estimates
of $T_{\rm eff}$ and $v \sin i$ from Section~\ref{sec:spectroscopy},
we have no direct information on the other fundamental physical
properties of the tertiary. Based on our spectroscopically measured
flux ratio of $L_3/L_1 = 0.108 \pm 0.012$ at 5187~\AA\ and the above
best-fit MIST isochrone, we infer $M_3 \approx 0.87~M_{\sun}$, $R_3
\approx 0.80~R_{\sun}$, and $T_{\rm eff} \approx 5490$~K. The
temperature is consistent with that estimated directly from our
spectra, and the radius ratio $R_3/R_1 \approx 0.54$ is not far from
the value we found in a different way at the end of
Section~\ref{sec:analysis}.

\section{Discussion}
\label{sec:discussion}

The $\sim$4\% discrepancy between the measured and predicted radius
for the K2 secondary in \vstar\ is in line with similar anomalies
displayed by other late-type stars having significant levels of
activity. While we do not detect any temperature suppression that
often accompanies radius inflation, the fractional effect in $T_{\rm
  eff}$ seen in other cases is typically half that of radius
inflation, or only about 100~K in this case, which is smaller than our
formal uncertainty.

\vstar\ is an X-ray source listed in the ROSAT All-Sky Survey
\citep{Voges:1999}, and is also reported to have shown at least one
X-ray flare during those observations \citep{Fuhrmeister:2003}. This
is a clear indication of magnetic activity in the system, though in
principle the source could be any of the three stars, or even all
three. From the ROSAT count rate of $0.082 \pm 0.012$ counts~s$^{-1}$
and the measured hardness ratio (${\rm HR1} = -0.22 \pm 0.14$) we
infer an X-ray flux of $F_{\rm X} = 5.9 \times
10^{-13}$~erg~cm$^{-2}$~s$^{-1}$, adopting the energy conversion
factor recommended by \cite{Fleming:1995}.  Using our distance
estimate of 90~pc we then derive an X-ray luminosity of $L_{\rm X} =
5.7 \times 10^{29}$~erg~s$^{-1}$.

While fairly common in late-type objects (particularly if rotating
rapidly), X-ray emission in stars much earlier than mid-F is generally
not easy to explain because they lack sufficiently deep surface
convective zones that are typically associated with magnetic activity
generated by the dynamo effect. For this reason, X-rays in these stars
are most often attributed to an unseen late-type companion
\citep[e.g.,][and references therein]{Schroder:2007}, which can easily
be hidden in the glare of the primary. Other mechanisms intrinsic to
earlier-type stars are possible, such as shocks and instabilities in
the radiatively driven winds, although these are not thought to be
able to explain variability such as the X-ray flaring mentioned above
\citep[see, e.g.][]{Schmitt:2004, Balona:2012}. We cannot completely
rule out a priori that the primary in \vstar\ is the main source of
the X-rays, but its much thinner convective envelope makes this seem
far less likely than an origin in a later-type star such as the
secondary or tertiary. Indeed, the MIST models indicate that the mass
of the convective envelope of the secondary is about 7.2\% of its
total mass, and that of the tertiary is 4.5\% (the value for the Sun
is 1.6\%), whereas the fractional mass of the primary's envelope is
only $3 \times 10^{-4}$.

The tertiary component in \vstar\ is a possible source for the X-rays,
if it were a rapidly rotating star. However, our spectroscopy suggests
it is not a fast rotator: we measure $v \sin i <
2$~\kms\ (Section~\ref{sec:spectroscopy}), although the projection
factor is unknown so it is concievable the equatorial rotation is much
faster. To estimate the true rotation period we used the age of the
system (2.2~Gyr) along with the gyrochronology relations of
\cite{Epstein:2014} and the estimated $B-V$ color of the star from the
MIST isochrones, and inferred $P_{\rm rot} \approx 18$~days. If
attributed entirely to the tertiary, the measured X-ray luminosity of
\vstar\ would be far in excess (by about an order of magnitude) of
what is expected for a star of this mass and rotation period,
according to studies of the relationship between stellar activity and
rotation \citep[e.g.,][]{Pizzolato:2003}. This argues the X-rays are
unlikely to originate mainly in the tertiary, although it is possible
it has some small contribution.

We are thus left with the secondary as the most probable site of the
bulk of the X-ray emission in \vstar. With the bolometric luminosity
given in Table~\ref{tab:dimensions} we compute $\log L_{\rm X}/L_{\rm
  bol} = -3.31$, a value that is close to the saturation level seen in
very active stars. The study of \cite{Pizzolato:2003} indicates that
this is in fact a typical value for a star of this mass with a
rotation period of 2.63~days, supporting our conclusion that the
secondary is the active star in the system. If that is the case, this
provides a natural explanation for its inflated radius.

Recent stellar evolution models that incorporate the effects of
magnetic fields have had some success in explaining radius inflation
in stars like the secondary \citep[see, e.g.,][]{Feiden:2012,
  Feiden:2013}. To achieve this, those models introduce a tunable
parameter that is the average strength of the surface magnetic field,
$\langle Bf\rangle$, where $B$ is the photospheric magnetic field
strength and $f$ the filling factor.  Measurements of magnetic field
strengths are very difficult to make in binary systems, let alone in
triple-lined systems such as \vstar, but they are essential in order
to validate the fits that these models provide.

A rough estimate of $\langle Bf\rangle$ for the secondary may be
obtained by taking advantage of a power-law relationship shown by
\cite{Saar:2001} to exist between $\langle Bf\rangle$ and the Rossby
number, $Ro \equiv P_{\rm rot}/\tau_c$, where $\tau_c$ is the
convective turnover time. For consistency with the work of
\cite{Saar:2001}, we take $\tau_c$ from the theoretical calculations
by \cite{Gilliland:1986}, which give $\tau_c \approx 29$~days for a
star with a temperature of 5020~K. The resulting Rossby number, $Ro
\approx 0.091$, together with the relation by \cite{Saar:2001} then
yields $\langle Bf\rangle \approx 1.1$~kG.\footnote{The same
  calculation applied to the tertiary star gives $\langle Bf\rangle
  \approx 70$~G, which is small compared to the secondary and supports
  the notion that it is not a very active star. The parameters for the
  primary star are outside of the range of validity of the
  \cite{Saar:2001} relation, but point to a magnetic field strength of
  only a few Gauss, again suggesting a very low activity level if the
  sustaining mechanism is the same as in late-type stars.}  An
independent way of estimating the magnetic field strength makes use of
the X-ray luminosity and the empirical relationship between that
quantity and the total unsigned surface magnetic flux, $\Phi = 4\pi
R^2 \langle Bf\rangle$. \cite{Pevtsov:2003} have shown in a study of
magnetic field observations of the Sun and active stars that the
relation holds over many orders of magnitude. With an updated version
of that relation by \cite{Feiden:2013}, and the measured radius of the
secondary, we obtain $\langle Bf\rangle \approx 1.0$~kG, which is
similar to our previous result. A magnetic field strength of this
order is quite consistent with values measured in many other cool,
active single stars \citep[see, e.g.,][]{Cranmer:2011, Reiners:2012}.

Our estimate of $\langle Bf\rangle \approx 1.0$~kG can serve as an
input to stellar evolution calculations that model the effects of
magnetic fields, and test their ability to match the measured size of
the secondary.

\vstar\ is attended by a distant third star that is physically bound:
we have shown that it has a similar radial velocity as the eclipsing
pair, a brightness perfectly consistent with that expected for a star
of its temperature at the same distance as the binary, and a motion on
the plane of the sky that is incompatible with a background object but
consistent with orbital motion in a highly inclined orbit around the
binary (possibly even coplanar with it). The system is thus a
hierarchical triple, which is not surprising given that
\citep{Tokovinin:2006} have shown that up to 96\% of all solar-type
binaries with periods shorter than 3 days have third components.

\section{Conclusions}
\label{sec:conclusions}

Our spectroscopic observations together with existing $uvby$
photometry have enabled us to derive accurate absolute masses and
radii for the eclipsing components good to better than 1.5\%, despite
the faintness of the secondary (only 3.6\% of the brightness of the
primary). \vstar\ thus joins the ranks of binary systems with the best
determined properties \citep[see][]{Torresetal:2010}. The highly
unequal masses provide increased leverage for the comparison with
stellar evolution models, and we find that the K2 secondary is about
4\% larger than predicted for its mass, though its temperature appears
normal. Thus, the star appears overluminous. The detection of the
system as an X-ray source is evidence of activity, and we have argued
that the source is the secondary component. This would provide at
least a qualitative explanation for the radius anomaly, which is also
seen in many other active stars with convective envelopes. We would
expect the secondary to have significant spot coverage, but the star
is much too faint compared to the primary for this to produce a
visible effect on the light curves.

\vstar\ is a good test case for recent stellar evolution models that
attempt to explain radius inflation in a more quantitative way by
including the effects of magnetic fields. To this end, we have
provided an estimate of the strength of the surface magnetic field on
the secondary ($\sim$1~kG).

Finally, we note that the study of this system would benefit from a
detailed chemical analysis of the primary star based on
high-resolution spectroscopy with broader wavelength coverage than the
45~\AA\ afforded by the material at our disposal.  This would remove
the metallicity as a free parameter in the comparison with stellar
evolution models, strengthening the results.

\vskip 10pt

\noindent{\bf Note added in proof:} A high-resolution ($R \approx
44,000$) echelle spectrum of \vstar\ with a signal-to-noise ratio of
220 in the \ion{Mg}{1}\,b region was obtained recently at the
Tillinghast reflector during the second quadrature (HJD 2,458,029.6,
phase 0.73). It shows no sign of activity (e.g., \ion{Ca}{2} H and K
or H$\alpha$ emission) in the brighter primary, supporting our
contention that this star is not particularly active.

\acknowledgments

We are grateful to P.\ Berlind, M.\ Calkins, D.\ W.\ Latham,
R.\ P.\ Stefanik, and J.\ Zajac for help in obtaining the
spectroscopic observations of \vstar, and to R.\ J.\ Davis and
J.\ Mink for maintaining the CfA echelle database over the years. We
also thank J.\ Choi for assistance in calculating the extent of
stellar envelopes, and the anonymous referee for helpful comments. We
acknowledge support from the SAO Research Experience for
Undergraduates (REU) program, which is funded by the National Science
Foundation (NSF) REU and Department of Defense ASSURE programs under
NSF grant AST-1659473, and by the Smithsonian Institution.
G.T.\ acknowledges partial support for this work from NSF grant
AST-1509375. This research has made use of the SIMBAD and VizieR
databases, operated at CDS, Strasbourg, France, and of NASA's
Astrophysics Data System Abstract Service.


\end{document}